\documentclass[final,5p,times,twocolumn]{elsarticle}

\usepackage{hyperref}
\usepackage{amssymb}
\usepackage{subfigure}
\usepackage{epsfig}
\usepackage{epstopdf}
\usepackage{algorithm}
\usepackage{xcolor}
\usepackage{graphicx}
\usepackage{amsmath}
\usepackage{algorithm}
\usepackage[noend]{algpseudocode}

\usepackage{lineno}

\biboptions{authoryear}

\makeatletter
\def\BState{\State\hskip-\ALG@thistlm}
\makeatother

\begin{document}

\begin{frontmatter}



\title{Basic Model of Purposeful Kinesis}


\author[LeicMath,NN]{A.N. Gorban\corref{cor1}}
\ead{a.n.gorban@le.ac.uk}
\author[LeicMath]{N. \c{C}abuko\v{g}lu}
\ead{nc243@le.ac.uk}

\address[LeicMath]{Department of Mathematics, University of Leicester, Leicester, LE1 7RH, UK}
\address[NN]{Lobachevsky University, Nizhni Novgorod, Russia}
\cortext[cor1]{Corresponding author}

\begin{abstract}
The notions of taxis and kinesis are introduced and used to describe two types of behaviour of an organism in  non-uniform conditions: (i) Taxis means the guided movement to more favorable conditions; (ii) Kinesis is the non-directional change in space motion in response to the change of conditions. Migration and dispersal of animals has evolved under control of natural selection. In a simple formalisation, the strategy of dispersal should increase Darwinian fitness. We introduce new models of purposeful kinesis  with diffusion coefficient dependent on fitness. The local and instant evaluation of Darwinian fitness is used, the reproduction coefficient. New  models include one additional parameter, intensity of kinesis, and may be considered as the {\em minimal models of purposeful kinesis}. The properties of models are explored by a series of numerical experiments. It is demonstrated how kinesis could be beneficial for assimilation of patches of food or of periodic fluctuations. Kinesis based on local and instant estimations of fitness is not always beneficial: for species with the Allee effect it can delay invasion and spreading. It is proven that kinesis cannot modify stability of positive homogeneous steady states.
\end{abstract}


\end{frontmatter}

\section{Introduction}

The notions of taxis and kinesis are introduced and used to describe two types of behaviour of an organism in  non-uniform conditions:
\begin{itemize}
\item Taxis means the guided movement to more favorable conditions.
\item Kinesis is the non-directional change in space motion in response to the change of conditions.
\end{itemize}
In reality, we cannot expect pure taxis without any sign of kinesis. On the other hand, kinesis can be considered as a reaction to the local change of conditions without any global information about distant sites or concentration gradients. If the information available to an organisms is completely local then taxis is impossible and  kinesis remains the only possibility of purposeful change of spatial behaviour in answer to the change of conditions. The interrelations between taxis and kinesis may be non-trivial: for example, kinesis can facilitate exploration and help to find non-local information about the living conditions. With this non-local information taxis is possible.

In this paper, we aim to present and explore a simple but basic model of purposeful kinesis. Kinesis is a phenomenon observed in a wide variety of organisms, down to the bacterial scale. {\em Purposeful} seems to imply a sort of intentionality that these organisms are incapable of. The terms `purpose' and `purposeful' are used  in mathematical modelling of biological
phenomena in a wider sense than in psychology. `Purpose' appears in a model when it includes optimisation.  The general concept of purposeful behaviour \citep{Rosenblueth1950} of animals requires the idea of evolutionary optimality \citep{Parker1990}. In many cases this optimality can be deduced from kinetic equations in a form of maximization of the average in time reproduction coefficient -- Darwinian fitness \citep{Metz1992, Gorban(1984), Gorban2007}. Application of this idea to optimization of behaviour is the essence of evolutionary game theory and its applications to population dynamics \citep{Hofbauer1998}.

There are three crucial questions for creation of an evolutionary game model:
\begin{enumerate}
\item Which information is available and usable? \cite{Dall2005} proposed a quantitative theoretical framework in evolutionary ecology for analysing the use of information by animal. Nevertheless, the question about information which can be recognised, collected and used by an animal requires empirical answers. Answering this question may be very complicated for analysis of taxis, which involves various forms of non-local information. For kinesis the situation is much simpler: the point-wise values of several fields (concentrations or densities) are assumed to be known \citep{Sadovskiy2009}.
\item What is the set of the available behaviour strategies? All the organisms, from bacteria to humans have their own set of available behaviour strategies, and no organism can be omnipotent. It is necessary to describe constructively the repertoire of potentially possible behaviours.
\item What are the statistical characteristics of the environment, in particular,  and what are the laws and correlations in the changing of environment in space and time? It is worth mentioning that all the changes in the environment should be measured by the corresponding changes of the reproduction coefficient.
\end{enumerate}

We use a {\em toy model} to illustrate the idea of purposeful kinesis. Assume that an animal can use one of two locations for reproduction. Let the environment in these locations can be in one of two states during the reproduction period, $A$ or $B$. The number of survived descendants is $r_A$ in state $A$ and $r_B$ in state $B$. After that, the offsprings leave the area of reproduction and their further survival does not depend on this area. Assume also that the change of states can be described by a Markov chain with transition probabilities $P_{A\to B}=p$ and $P_{B\to A}=q$. These assumptions answer Question 3.

The animal is assumed to be very simple: it can just evaluate the previous state of the location where it is now but cannot predict the future state. There is no memory: it does not remember the properties of the locations where it was before. This is the answer to question 1.

Finally, there is only one available behaviour strategy: to select the current (somehow chosen)  location or to move to another one. There exists resources for one jump only and no `oscillating' jumps between locations are possible. This means that after the change of location the animal selects the new location for reproduction independently of its state. Thus Question 2 is answered.

Analysis of the model is also simple. If the state of the location is unknown then the probability of finding it in state $A$ is $\frac{q}{p+q}$ and the probability of finding it in state $B$ is $\frac{p}{p+q}$; these are the stationary probabilities of the Markov chain. The expectation of the number of offsprings without arbitrary information is
$$r_0=\frac{qr_A+pr_B}{p+q}.$$
If an animal chooses for reproduction the location with the previous state $A$ then the conditional expectation of the number of offsprings is
$r\left|_A\right.=(1-p) r_A+ pr_B$. If it chooses the location with the previous state $B$ then the expected number of offsprings is $r\left|_B\right.=(1-q) r_B+ qr_A$.

If the animal is situated in the location  with the previous state $A$, and $r\left|_A\right.<r_0$, then the change of location will increase the expected number of offsprings. Analogously, if it is situated  in the location with the previous state $B$, and $r\left|_B\right.<r_0$, then the change of location will increase the number of offsprings.

We have obtained the simplest model with mobility dependent conditionally expected reproduction coefficient $r\left|_{\bullet}\right.$ under given local conditions: if $r\left|_{\bullet}\right.$ is less than the value $r_0$ expected  for the indefinite situation then jump, else stay in the same location. This is the essence of purposeful kinesis for this toy model.

It is very difficult to find  realistic space and time correlations in the environment during the evolution of animals under consideration.
The answers to Questions 1 and 2 for real animals are also non-obvious, but the main idea can be utilised for the modelling of kinesis. We expect that the dynamics of the models could provide insight, regardless of whether parameters were obtained from optimization of real Darwininan fitness or just the structure of equations was guessed on the basis of this optimization.

In this paper, we study PDE models of space distribution. We start from the classical family of models. \cite{Patlak1953}, and  \cite{Keller1971} proposed a PDE system which is widely used for taxis modelling \citep{Hillen2009}.

\begin{equation}
\begin{split}
&\partial_t u( t,x) =\nabla\left(k_1(u,s) \nabla u  + k_2(u,s) u \nabla s \right) + k_3(u,s)u, \\
&\partial_t s ( t,x) = D_s \nabla^2 s +k_4(u,s) - k_5(u,s)s,
\end{split}
\label{KSModel}
 \end{equation}
where
\begin{itemize}
\item[] $u\geq 0$ is the population density,
\item[] $s\geq 0$ is the concentration of the attractant,
\item[] $D_s\geq 0$ is the  diffusion coefficient of  the attractant,
\item[] coefficients $k_i(u,s)\geq 0$.
\end{itemize}

Coefficient $k_1(u,s)$ is a diffusion coefficient of the animals. It depends on the population density $u$ and on the concentration of the attractant $s$. Coefficient $k_2(u,s)$ describes intensity of population drift.

Special random processes were introduced for `microscopic' theory of dispersal in biological systems by \cite{Othmer1988}. They consist of two modes: (i) position jump or kangaroo processes, and (ii) velocity jump processes:
\begin{itemize}
\item The kangaroo process comprises a sequence of pauses and jumps. The distributions of the waiting time, the direction and distance of a jump are fixed;
\item The velocity jump process consists of a sequence of `runs' separated by reorientations, after which a new velocity is chosen.
\end{itemize}

Equations (\ref{KSModel}) can be produced from kinetic (transport) models of velocity--jump random processes \citep{Othmer2000, Othmer2002} in the limit of large number  of animals and small density gradients under an appropriate scaling of space and time. The higher approximations are also available in the spirit of the Chapman--Enskog expansion from physical kinetics \citep{Chapman1970}. \cite{Chalub2004} found sufficient conditions of absence of finite-time-blow-ups in chemotaxis models. \cite{Turchin1989} demonstrated that attraction (and repulsion) between animals could modify the space dispersal of population if this interaction is strong enough. \cite{Mendez2012} derived reaction-dispersal-aggregation equations from Markovian reaction-random walks with density-dependent transition probabilities. They have obtained a general threshold value for dispersal stability and found the sufficient conditions for the emergence of non-trivial spatial patterns.
\cite{Grunbaum1999} studied how the advection-diffusion equation can be produced for organisms (``searchers'') with different food searching strategies with various {\em turning rate} and {\em turning time} distributions, which depend on the density of observed food distribution.

The family of models (\ref{KSModel}) is rich enough and the term $\nabla(k_1(u,s) \nabla u)$ can be responsible for modelling of kinesis: it describes non-directional motion in space with the {\em diffusion coefficient} $D=k_1(u,s) $. This coefficient depends on the local situation represented by $u$ and $s$. In some sense, the family of models (\ref{KSModel})  is even too rich: it includes five unknown functions $k_i$ with the only requirement, the non-negativity.

\cite{Cosner2014} reviewed PDE reaction--advection--diffusion models for the ecological effects and evolution of dispersal, and mathematical methods for analyzing those models. In particular, he discussed a series of optimality or evolutionary questions which arose naturally: Is it better for the predators to track the prey density, the prey's resources, or some kind of combination? Is it more effective for predators to slow down their random movement when prey are present or to use directed movement up the gradient of prey density? Should either predators or prey avoid crowding by their own species? \cite{Cosner2014} presented also examples when diffusion is harmful for the existence of species: if the average in space of the reproduction coefficient is negative for all distributions of species then for high diffusion there is no steady state with positive total population even if there exist steady states with positive total population for zero or small  diffusion (for connected areas). The possibility of organisms moving sub- or super-diffusively, e.g. L\'evy walks, fractional diffusion, etc. (see, for example the works by \cite{Chen2010, Mendez2010}), can be combined with the idea of purposeful mobility (see, for example the works by \cite{Chen2010, Mendez2010}) but  we limit analysis in this paper by the classical PDE.

In this work, we study the population dispersal without taxis, therefore, the advection coefficients $k_2$ is set below to zero.
Such dispersal strategy seems to be quite limited comparing to the general kinesis+taxis dispersal system. Nevertheless, \cite{Nolting2015} demonstrated on the jump models that  the purely kinesis (non-directional dispersal strategy) allows foragers to identify efficiently intensive search zones without taxis and are more robust to changes in resource distribution.

We also assume strong connection between the {\em reproduction coefficient} $r=k_3$ and the diffusion coefficient. The reproduction coefficient characterises both the competitive abilities of individuals, and their  fecundity. The Darwinian fitness  is the average reproduction coefficient in a series of generations \citep{Haldane1932, Metz1992, Gorban2007}. Dynamics maximise the Darwinian fitness of survivors (this is formalisation of natural selection). Unfortunately, evaluation of this quantity is non-local in time and requires some knowledge of the future. Therefore, we use below the local in time and space estimation of fitness and measure the well-being by the instant and local value of  the reproduction coefficient $r$. This is a rather usual approach but it should be used with caution: in some cases, the optimisation of the local criteria can worsen  the long-time performance. We describe one such situation below: use a locally optimised strategy of kinesis may delay invasion and spreading of species with the Allee effect. On another hand, we demonstrate how kinesis controlled by the local reproduction coefficient may be beneficial for assimilation of patches of food or periodic fluctuations.

\section{Main Results}

\subsection{The ``Let well enough alone" model}

The kinesis strategy controlled by the locally and instantly evaluated well-being can be described in simple words: Animals stay longer in good conditions and leave quicker bad conditions. If the well-being is measured by the instant and local reproduction coefficient then the minimal model of kinesis can be written as follows:
\begin{equation}\label{KinesisModel}
\boxed{
 \partial_t u_ i ( x,t)
   =  D_{0i} \nabla \left(e ^{-\alpha_i r_i(u_1,\ldots,u_k,s)} \nabla u_i \right)+r_i (u_1,\ldots,u_k,s) u_i,
 }
   \end{equation}
   where:
\begin{itemize}
\item[] $u_i$ is the population density of  $i$th species,
\item[]   $s$ represents the abiotic characteristics of the living conditions (can be multidimensional),
\item[] $r_i$ is the reproduction coefficient, which depends on all $u_i$ and on  $s$,
\item[] $D_{0i}>0$ is the equilibrium diffusion coefficient (defined for $r_i=0$),
\item[]   The coefficient $\alpha_i>0$  characterises dependence of the diffusion coefficient on the reproduction coefficient.
\end{itemize}
Equations (\ref{KinesisModel}) describe dynamics of the population densities for arbitrary dynamics of $s$. For the complete model the equations for environment $s$ should be added.
The space distribution strategy is summarised in the diffusion coefficient $D_i=D_{0i}e ^{-\alpha_i r_i} $, which depends only on the local in space and time value of the reproduction coefficient. Diffusion depends on well-being measured by this coefficient.
We can see that the new models add one new parameter per species to the equations (instead of function $k_1(u,s)$ in (\ref{KSModel})). This is the kinesis constant $\alpha_i$. It can be defined as

$$\alpha_i=-\frac{1}{D_{0i}}\left.\frac{d D_i(r_i)}{d r_i}\right|_{r_i=0}.$$
In the first approximation, $D_i=D_{0i}(1-\alpha_ir_i)$. The exponential form in (\ref{KinesisModel}) guarantees positivity of the coefficient $D_i$ for all values of $r_i$.

For good conditions ($r_i>0$) diffusion is slower than at equilibrium ($r_i=0$) and for worse conditions ($r_i<0$) it is faster. Equations (\ref{KinesisModel}) just formalise a simple wisdom: do not change the location that is already good enough  ({\em let well enough alone}) and run away from bad location.

We analyse below how the dependence of diffusion on well-being effects patch dynamics and waves in population dynamics.

\subsection{Stability  of uniform distribution}

The positive uniform steady state $(u^*,s^*)$ satisfies the equation: $r_i (u^*_1,\ldots,u^*_k,s^*)=0$.

The linearised equations near the positive uniform steady state are
\begin{equation}
\partial_t \delta u_ i ( x,t)
   =d_i \nabla^2 (\delta u_i)+u_i \left(\sum_j r_{i,j}\delta u_j+r_{i,0}\delta s\right)  ,
\end{equation}
   where
\begin{itemize}
\item[] $\delta u_i$ is the deviation  of the population density of  $i$th animal from equilibrium $u^*$, $\delta s=s-s^*$,
\item[]  $r_{i,j}=\partial r_i/\partial u_j|_{(u^*,s^*)}$,  $r_{i,0}=\partial r_i/\partial s|_{(u^*,s^*)}$ are  derivatives of $r$ at equilibrium.
\end{itemize}

These linearised equation are the same as for the system without kinesis (with constant diffusion coefficients).
Therefore, {\em kinesis does not change stability of positive uniform steady states}. Moreover, near such a steady state linearised equations for a system with kinesis are the same as for the system with constant diffusion coefficient.

There is an important difference between possible  dynamic consequences of taxis and kinesis: we proved that kinesis cannot modify stability of homogeneous steady states, whereas \cite{Tyutyunov2017} demonstrated that taxis  can destabilise them.

\subsection{Utilisation of a patch of food}

\begin{figure}[ht]
\centering\includegraphics[width=0.45\textwidth]{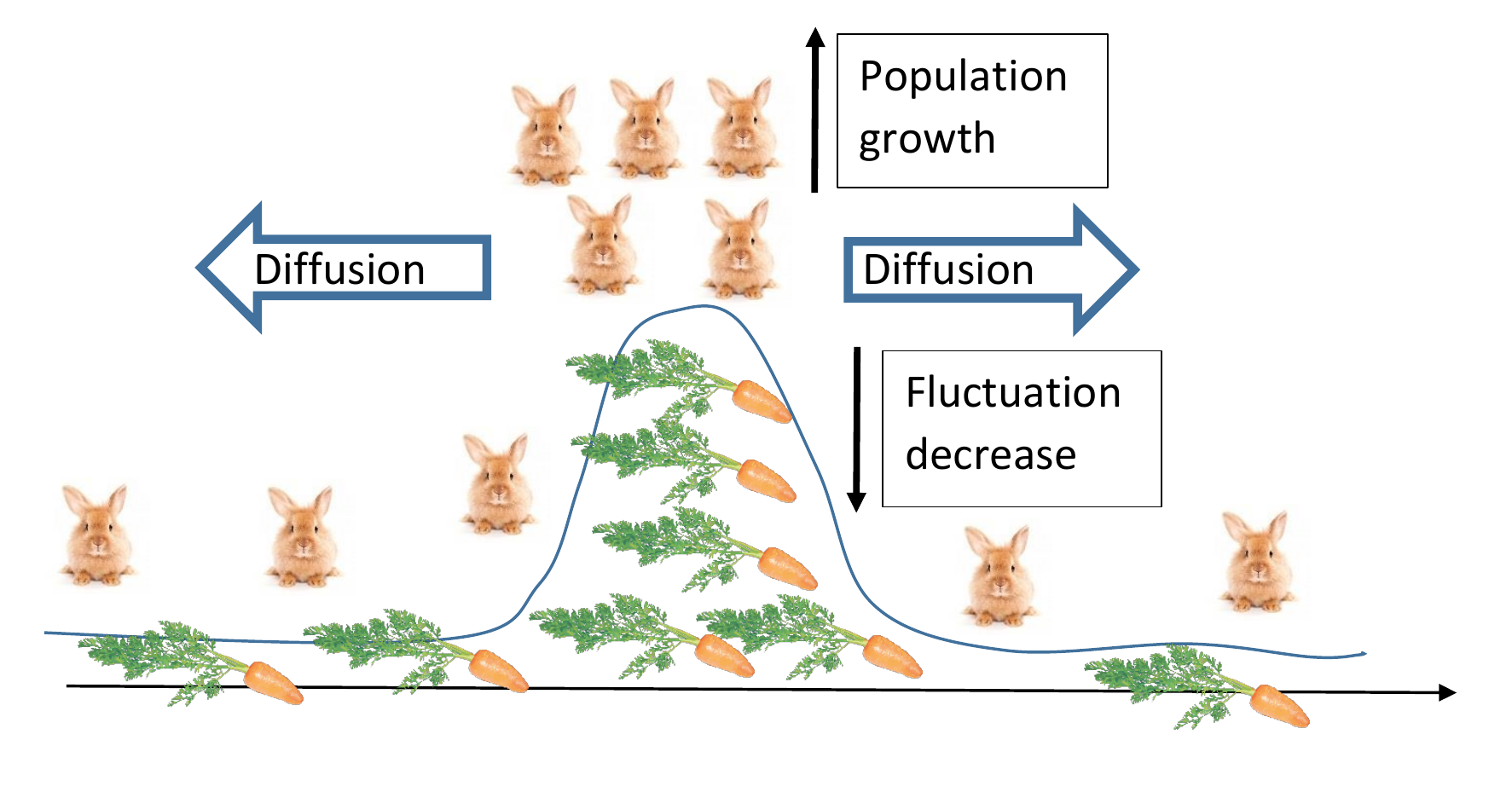}
\caption{A schematic representation of a patch of food.\label{Fig1:Rabbits}}
\end{figure}

As a first test for the new model we used utilisation of a patch of food (a sketch of this gedankenexperiment is presented in Fig.~\ref{Fig1:Rabbits}). Concentration of food in patches is one of the standard ecological situations. \cite{Nonaka2007} considered ``clumpiness''  as a main characteristic of the food distribution and developed an agent-based model for analysis of optimal foraging.
\cite{Grunbaum1998} studied foraging in population of the ladybird beetle ({\em Coccinella septempunctata}), while preying on the goldenrod aphid ({\em Uroleucon nigrotuberculatum}). He used both experimental observation and PDE models and analysed nonuniform ``aggregated'' distributions, in which foragers accumulate at resource concentrations, and evaluated parameters of foragers' strategy  from experimental data.

From our point of view, the potential of PDE models is not exhausted despite of growing popularity of the multiagent models in dynamics of space distribution of populations \citep{Rahmandad2008}. Let us compare two models:
\begin{itemize}
\item{  A system of one PDE for population  with kinesis and one ODE for substrate:
\begin{equation}
\begin{split}
&\partial_t u ( t,x) = D \nabla\left(e^{-\alpha ( a s( t,x ) -b ) } \nabla u \right) + ( a s( t,x) -b) u( t,x), \\
&\partial_t s ( t,x) = -g u( t,x) s( t,x )+d;
\end{split}
\label{Model1}
 \end{equation}
 }
\item{ A system of one PDE for population with the constant diffusion coefficient (i.e. without kinesis) and the same ODE for substrate:
\begin{equation}
\begin{split}
&\partial_t u( t,x) = D \nabla^2 u  + ( a s( t,x) -b ) u( t,x), \\
&\partial_t s( t,x) = -g u( t,x) s( t,x)+d.
\end{split}
\label{Model2}
 \end{equation}
 }
\end{itemize}
These models are particular realisations of the system (\ref{KSModel})

For the computations experiment, to solve partial differential equations, first MATLAB \cite{pdpe2017} function has been used for space dimension one. For two-dimensional results below, the MATHEMATICA \cite{NDSolve2014} solver with Hermite method and Newton's divided difference formula has been used.

We selected 1D benchmark (Fig.~\ref{Fig1:Rabbits}, compare to Fig.~1 in work of \cite{Grunbaum1998}) on the interval $[-50,50]$   with boundary conditions and with the initial conditions:
$$s( 0,x) = A e^{-\frac{x^2}{2}}, u( 0,x) =1, A=4.$$ The values of the constants are: $D=10$, $\alpha=5$, $a=2$, $b=1$, $g=1$, $d=1$.

\begin{figure}[ht]
\centering
\includegraphics[width=0.45\textwidth]{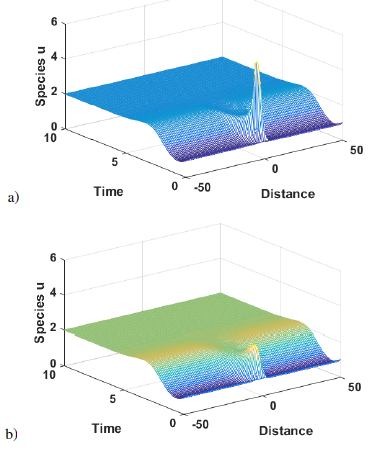}
\caption{Utilisation of a food patch. Population density burst and relaxation: a) for animals with kinesis and b) for animals 	without kinesis. \label{PatchDecay2D}}
\end{figure}

\begin{figure}[ht]
\centering
a)\includegraphics[width=0.45\textwidth]{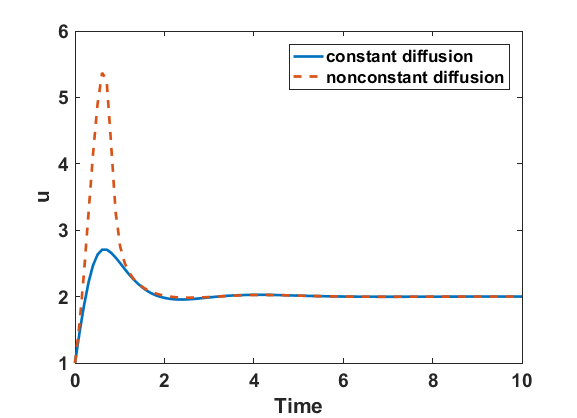}
b)\includegraphics[width=0.45\textwidth]{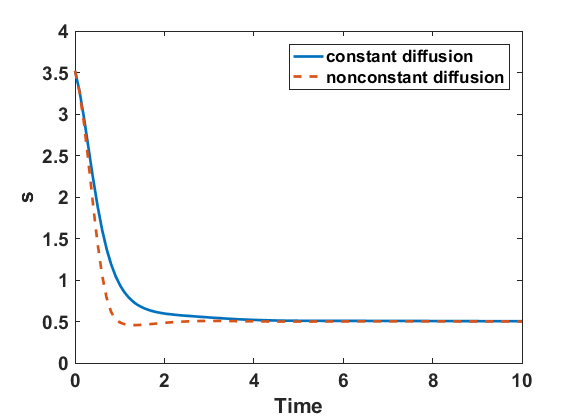}
\caption{Utilisation of a food patch: a) dynamics of population density at the centre of patch, b) dynamics of the food density at the centre of patch. \label{PachDecay1D}}
\end{figure}

It is the first expectation that the proper kinesis should improve the ability of animals to survive in a clumpy landscape. We can see from Figs.~\ref{PatchDecay2D}, \ref{PachDecay1D} that the density burst for the system with kinesis is higher and the utilisation of fluctuation of substrate goes faster than without kinesis.

The fluctuation of food decreases faster for the system with kinesis. The population density increases to a higher level for system with kinesis. This is essentially non-linear effect because in the linear approximation near uniform equilibrium models with kinesis (\ref{Model1})  and without kinesis (\ref{Model2} coincide.

\subsection{Utilisation of fluctuations in food density}

For the second benchmark we consider fluctuations of substrate, which are periodic in space and time. Our gedankenexperimet includes two populations of animals. The only difference between them is that the first population diffuses with kinesis (population density $v$), whereas the second (population density $u$) just diffuses with the constant diffusion coefficient (no kinesis). The equilibrium values of the diffusion coefficients coincide. These populations interact by consuming the same resource as it is described by equations \ref{Eq:fluct} below
\begin{equation}
\begin{split}\label{Eq:fluct}
&\partial_t u ( t,x) = D \nabla^2 u + (a s( t,x) -b) u( t,x);\\
&\partial_t v( t,x) = D \nabla \left(e^{-\alpha ( a s( t,x) -b) } \nabla v\right) + ( a s( t,x) -b)v( t,x);\\
&\partial_t s ( t,x) = -g(  u+v ) s + d[  1+\delta \sin( w_1 t)\sin( w_2 x)],
\end{split}
 \end{equation}
with zero-flux boundary conditions and with the initial conditions:
 $$s( 0,x) = 0.5, u( 0,x) =1, v( 0,x) =1.$$
 The values of constants are: $D=10$, $\alpha=5$, $a=2$, $b=1$, $g=1$, $d=1$, $w_1=w_2=1$.

\begin{figure}
\centering
\includegraphics[width=0.45\textwidth]{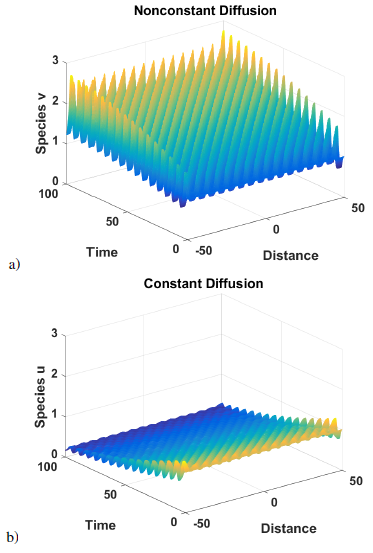}
\caption{Dynamics of population densities in fluctuating conditions: a) growth of subpopulation with kinesis, b) extinction of subpopulation without kinesis. \label{FluctDecay2D}}
\end{figure}

\begin{figure}
\centering\includegraphics[width=0.45\textwidth]{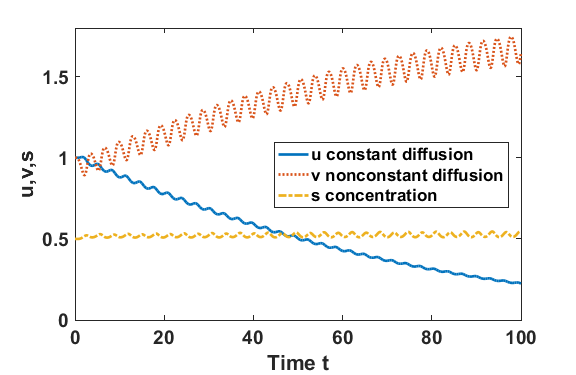}
\caption{Dynamics of population densities in fluctuating conditions at one point (the centre of the interval). Concurrent exclusion of the population without kinesis by the population with kinesis. \label{ExtinctionFluc}}
\end{figure}

Animals with kinesis have evolutionarily benefits in the explored non-stationary condition. We observe extinction of the population without kinesis (Figs.~\ref{FluctDecay2D} and \ref{ExtinctionFluc}). This is the concurrent exclusion of the animals without kinesis by the animals with kinesis. At the same time, the fluctuations of the population with kinesis in space and time are lager then for the population without it. This effect was expected: animals with kinesis rarely leave the beneficial conditions an jump more often from the worse conditions. In the conditions with the reproduction coefficient $r>0$ their density grows faster and in the worse condition (with $r<0$) it decreases faster than for animals with the constant diffusion coefficient.

\subsection{Spreading of a population with the Allee effect}

The reproduction coefficient of a population takes its maximal value at zero density and monotonically decays with the density growth in the simplest models of logistic growth and their closest generalisations. It is widely recognised that such a monotonicity is an oversimplification: The reproduction coefficient is not a monotonic function of the population density \citep{Allee1949, Odum1971}. This is the so-called {\em Allee effect}. The assumption of the negative growth rate  for small values of the population density is sometimes also included in the definition of the Allee effect.  The Allee effect is often linked to the low probability of finding a mate in a low density population but non-monotonicity of dependence of the reproduction coefficient on the population density and existence of the positive optimal density can have many different reasons. For example, any form of cooperation in combination with other density-dependent factors could also produce a non-monotonic reproduction coefficient and existence of optimal population density.

\begin{figure}[ht]
\centering\includegraphics[width=0.4\textwidth]{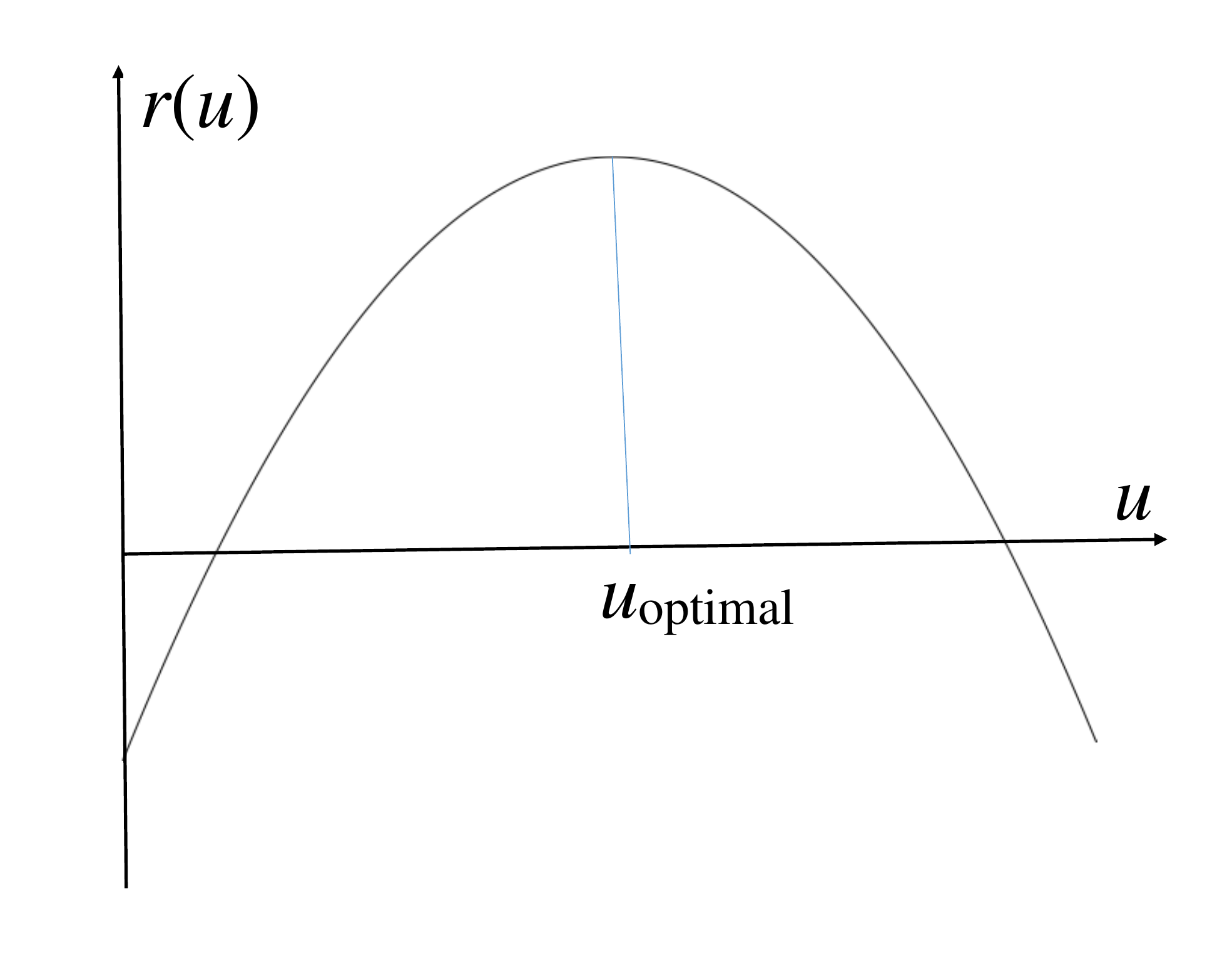}
\caption{Reproduction coefficient with the  Allee effect.\label{AlleyReproduction}}
\end{figure}

The simplest polynomial form of the reproduction coefficient with the Allee effect is $r(u)=r_0(K-u)( u-\beta)$. A typical dependence $r(u)$ with the Allee effect is presented in Fig.~\ref{AlleyReproduction}. The optimal density corresponds to the maximal value of $r$ (by definition). The evolutionarily optimal strategy for populations with the Allee effect is life in clumps with optimal density when the average density is lower than the optimal density \citep{Gorban1989}. This clumpiness appears even in homogeneous external conditions and is the most clear manifestation of  the Allee effect in ecology. There are multiple dynamical consequences of the  Allee effect \citep{McCarthy1997, Bazykin1998}. In combination with diffusion it leads to a possibility of  spread of invasive species via formation, interaction and movement of separate patches even in homogeneous external conditions \citep{Petrovskii2002, Morozov2006}.

The reaction--diffusion equations for a single population with  the Allee effect in dimensionless variables are below for a system without kinesis (\ref{AlleyWithout}) and  for a system with kinesis (\ref{AlleyWith}).

\begin{eqnarray}
\partial_t u ( t,x)& = &D \nabla ( \nabla u ) + ( 1-u)( u-\beta) u( t,x), \label{AlleyWithout}\\
\partial_t u( t,x)& = &D \nabla\left(e^{-\alpha ( 1-u) ( u-\beta ) } \nabla u \right)\\
&&\nonumber + ( 1-u )( u-\beta) u( t,x).\label{AlleyWith}
\end{eqnarray}
The values of the constants are: $D=1$, $\alpha=10$, $\beta=0.2$.

We study invasion of a small, highly concentrated population  into a homogeneous environment. Equations (\ref{AlleyWithout}) and   (\ref{AlleyWith}) are solved for one space variable $x\in[-50,50]$  with  zero-flux boundary  boundary conditions and with the initial conditions:
\begin{equation}\label{1dInitial}
u( 0,x) =A e^{-\frac{x^2}{2}}, \; A=1.
\end{equation}

\begin{figure}
\centering
\includegraphics[width=0.45\textwidth]{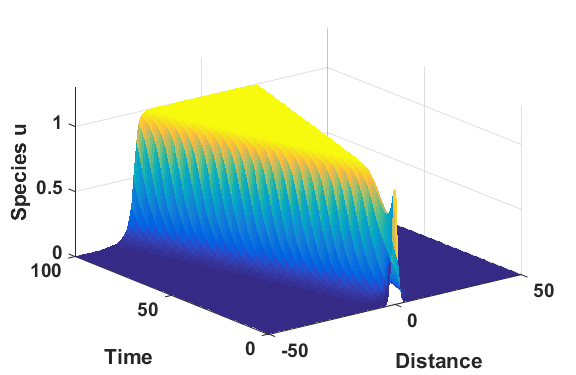}
b)\includegraphics[width=0.45\textwidth]{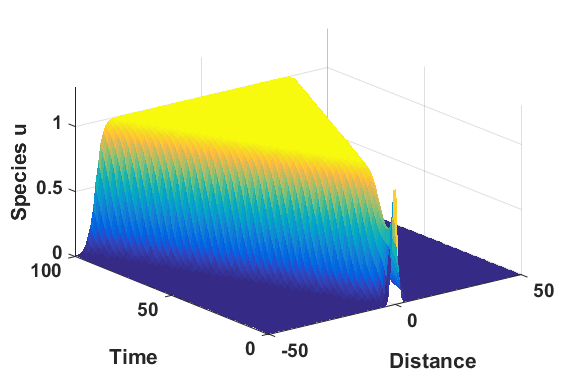}
\caption{Evolution of a small, highly concentrated population with the  Allee effect from the Gaussian initial conditions (\ref{1dInitial}) in a homogeneous environment: (a) for animals with kinesis, (b) for animals without kinesis (\ref{AlleyWithout}). The values of constants are: $D=1$, $\alpha=5$, $\beta=0.2$.  \label{AlleeSpreading2D}}
\end{figure}

\begin{figure}
\centering
a)\includegraphics[width=0.45\textwidth]{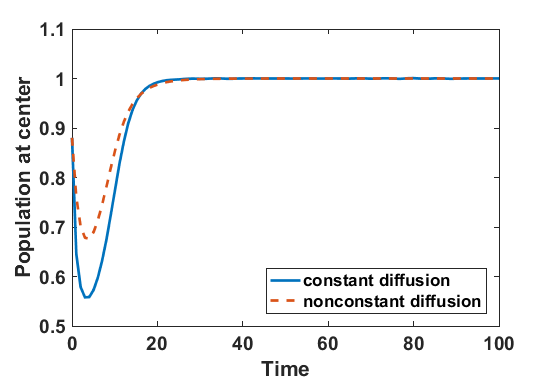}
b)\includegraphics[width=0.45\textwidth]{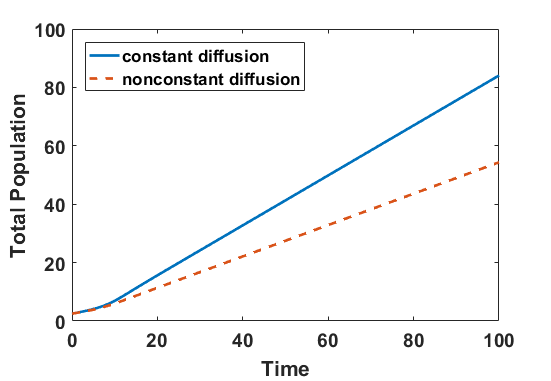}
\caption{Evolution of a small, highly concentrated population  with  the Allee effect from the Gaussian initial conditions (\ref{1dInitial}) in a homogeneous environment: (a) population density at the centre of the drop, (b) total population dynamics (numerically integrated over space). The values of constants are: $D=1$, $\alpha=5$, $\beta=0.2$. \label{AlleySpreading1D}}
\end{figure}

\begin{figure*}
\centering
\includegraphics[width=0.25\textwidth]{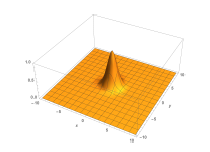}
\caption{Initial distribution ($t=0$): $u( 0,x) =A e^{-\frac{x^2+y^2}{2}}$, $A=1$.\label{2Dini}}
\end{figure*}

\begin{figure*}
\centering
{\includegraphics[width=0.7\textwidth]{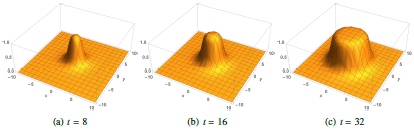}}
\caption[]{2D Allee effect {\em with kinesis} $D=0.2$, $\alpha=5$, $\beta=0.1$, with Gaussian initial distribution (Fig.~\ref{2Dini}). }
\label{2DkinesisLowDiff}
\end{figure*}

\begin{figure*}
\centering
{\includegraphics[width=0.7\textwidth]{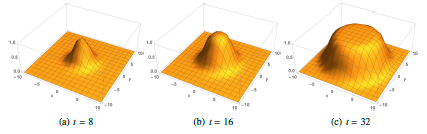}}
\caption[]{2D Allee effect {\em without kinesis} $D=0.2$, $\alpha=5$, $\beta=0.1$, with Gaussian initial distribution (Fig.~\ref{2Dini}). }
\label{2DNOkinesisLowDiff}
\end{figure*}

The results of the numerical experiments (Figs.~\ref{AlleeSpreading2D} and \ref{AlleySpreading1D}) demonstrate that kinesis may delay invasion and spreading of species with the  Allee effect. The width of the cluster grows faster for the system without kinesis (Fig.~\ref{AlleeSpreading2D}), and the total population dynamics numerically integrated over space (Fig.~\ref{AlleySpreading1D}) also demonstrates the faster growth of population without kinesis. The delay in spreading appears because the animals with kinesis rarely leave dense clusters, whereas animals without kinesis are spreading in areas with lower values of the reproduction coefficient and populate them. this effect is also reproduced in two-dimensional case presented in Figs.~\ref{2DkinesisLowDiff} and \ref{2DNOkinesisLowDiff}: an initial Gaussian drop (Fig.~\ref{2Dini}) grows with kinesis (Fig.~\ref{2DkinesisLowDiff}) slower than than without kinesis (Fig.~\ref{2DNOkinesisLowDiff}).

\begin{figure*}
\centering
{\includegraphics[width=0.7\textwidth]{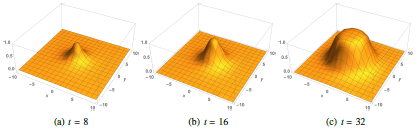}}
\caption[]{2D Allee effect {\em with kinesis}, $D=0.5$, $\alpha=5$, $\beta=0.1$, with Gaussian initial distribution (Fig.~\ref{2Dini}).}
\label{2DkinesisHighDiff}
\end{figure*}

\begin{figure*}
\centering
{\includegraphics[width=0.7\textwidth]{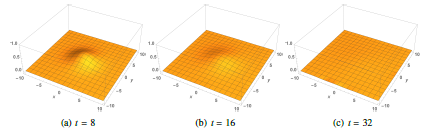}}
\caption[]{2D Allee effect {\em without kinesis} $D=0.5$, $\alpha=5$, $\beta=0.1$, with Gaussian initial distribution (Fig.~\ref{2Dini}). }
\label{2DNOkinesisHighDiff}
\end{figure*}

This effect of faster spreading could also lead to extinction for a population with Allee effect. For small population density $u<\beta$ the
reproduction coefficient is negative. If diffusion is so fast that the local concentration becomes lower than the threshold $\beta$ then the extinction of population follows. In Fig.~\ref{2DNOkinesisHighDiff} we can see how the population without kinesis vanishes for high diffusion, whereas the population with kinesis persists for the same diffusion (Fig.~\ref{2DkinesisHighDiff}) because it keeps low mobility at locations with high reproduction coefficient.


\section{Discussion}

\begin{figure}
\centering{
\includegraphics[width=0.43\textwidth]{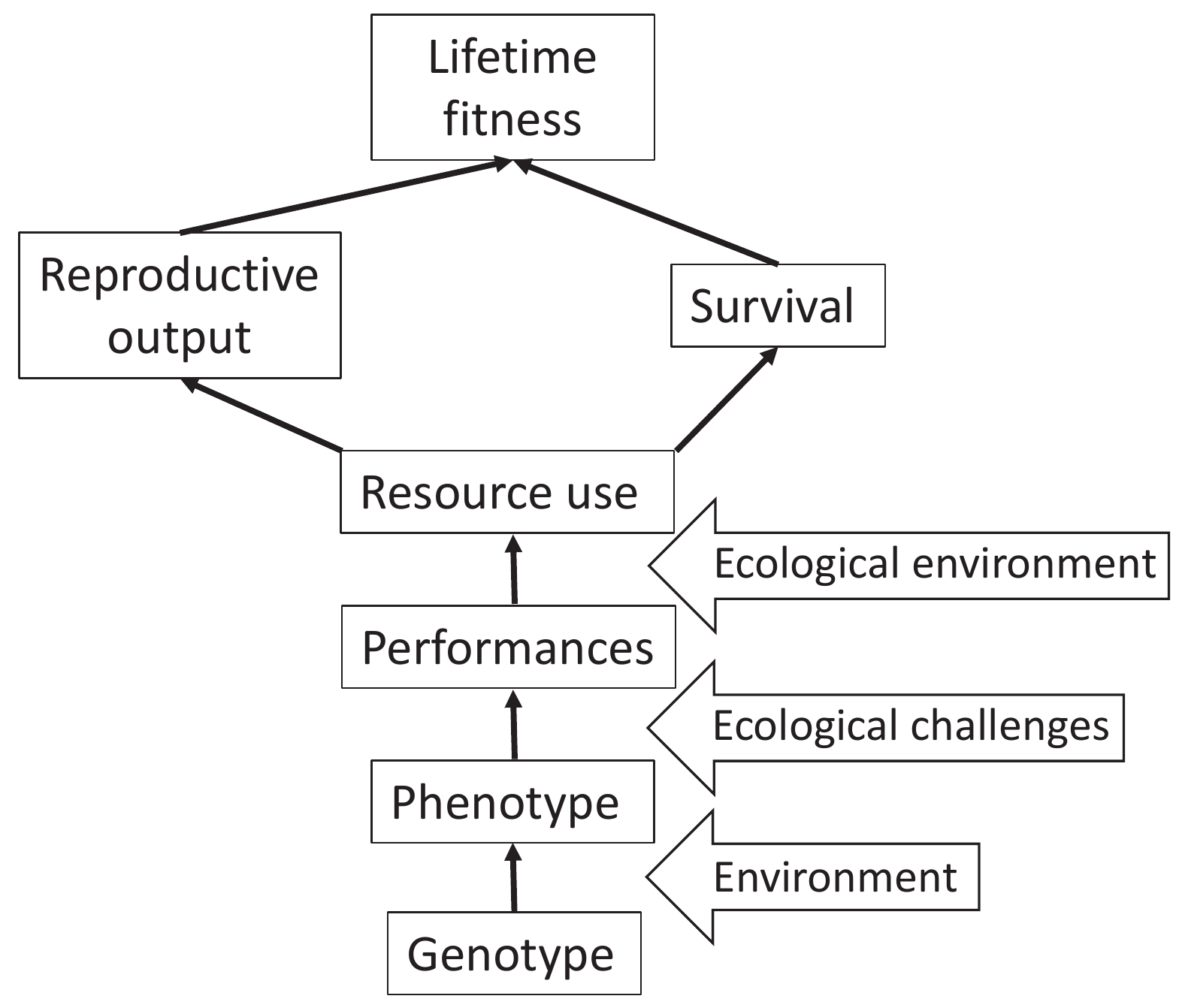}
}\caption{ Flow diagram showing the paths through from genotype to Darwinian fitness. Genotype in combination with environment determines the phenotype  up to some individual variations.  Phenotype determines  the limits of an individual's ability to perform day-to-day behavioural  answer to main ecological challenges (performances). Performance capacity interacts with the given ecological environment and determines the resource use, which is  the key internal factor determining reproductive output and survival.
 \label{PerfScheme}}
\end{figure}

We suggested a model of purposeful kinesis with the diffusion coefficient directly dependent on the reproduction coefficient. This model is a straightforward formalisation of the rule: ``Let well enough alone''. The well-being is measured by local and instant values of the reproduction coefficient. \cite{Gorban2016} have discussed the problems of definition of instant individual fitness in the context of physiological adaptation. Let us follow here this analysis in brief. The proper Darwinian fitness is defined by the long-time asymptotic of kinetics. It is non-local in time because it is the average reproduction coefficient in a series of generations and does not characterize an instant state of an individual organism \citep{Haldane1932, Maynard-Smith1982, Metz1992, Gorban2007}. The synthetic evolutionary approach starts with the analysis of genetic variation and studies the phenotypic effects  of that  variation  on physiology. Then it goes to the performance of organisms in the sequence of generations (with adequate analysis of the environment) and, finally, it has to return to Darwinian  fitness.
The ecologists and physiologists  are focused, first of all,  on  the  observation  of  variation  in individual performance \citep{Pough1989}. In this approach we have to measure the individual performance and then link it to the Darwinian fitness. This link is not obvious.  Moreover, the dependence between the individual performance and  the Darwinian fitness is not necessarily  monotone. (This observation was partially formalized in the theory of $r-$ and $K-$ selection \citep{MacArthurWilson1967, Pianka1970}.) The notion  `performance' in ecology is `task--de\-pen\-dent' \citep{Wainwright1994} and refers to an organism's ability to carry out specific behaviours and tasks: to capture prey, escape predation, obtain mates, etc. Direct instant measurement of Darwinian fitness is impossible but it is possible to measure various instant performances several times and treat them as the components of fitness in the chain of generations. The relations between performance and lifetime fitness are sketched on flow-chart (Fig.~\ref{PerfScheme}) following \cite{Wainwright1994} with minor changes. Darwinian fitness may be defined as the lifetime fitness averaged in a sequence of generations.

The instant individual fitness is the most local in time level in the multiscale hierarchy of measures of fitness: instant individual fitness $\to$ individual life fitness $\to$ Darwinian fitness in the chain of generations. The quantitative definition of the instant and local fitness is given by its place in the equations. The change of the basic equation will cause the change of the quantitative definition.

We have used the instant and local reproduction coefficient $r$ for defining of purposeful kinesis. The analysis of several benchmark situations demonstrates that, indeed, sometimes this formalisation works well. Assume that this coefficient $r(u,s)$ is a monotonically decreasing function of $u$ for every given $s$ and  monotonically increasing function of $s$ for any given $u$. Then our benchmarks give us the following hints:
\begin{itemize}
  \item It the food exists in low-level uniform background concentration and in rare (both in space and time) sporadic patches then  purposeful kinesis defined by the instant and local reproduction coefficient (\ref{KinesisModel}) is evolutionarily beneficial and allows animals to utilise the food patches more intensively (see Figs.~\ref{PatchDecay2D} and \ref{PachDecay1D});
  \item If there are periodic (or almost periodic) fluctuations in space and time of the food density $s$ then purposeful kinesis defined by the instant and local reproduction coefficient (\ref{KinesisModel}) is evolutionarily beneficial and allows animals to utilize these fluctuations more efficiently (see Figs.~\ref{FluctDecay2D} and \ref{ExtinctionFluc}).
  \item If the reproduction coefficient $r(u,s)$ is not a  monotonically decreasing function of $u$ for every given $s$ (the Allee effect) then the ``Let well enough alone" strategy may delay the spreading of population (see Figs.~\ref{AlleeSpreading2D}, \ref{AlleySpreading1D}, \ref{2DkinesisLowDiff}, and \ref{2DNOkinesisLowDiff}). This strategy can lead to the failure in the evolutionary game when the colonization of new territories is an important part of evolutionary success. This manifestation of the difference between the local optimisation and the long-time evolutionary optimality is important for understanding of the evolution of dispersal behaviour. At the same time, the ``Let well enough alone" strategy can prevent the effects of extinction caused by too fast diffusion (see Figs.~\ref{2DkinesisHighDiff} and \ref{2DNOkinesisHighDiff}) and, thus, decrease the effect of harmful diffusion described by \cite{Cosner2014}.
  \end{itemize}
These results of exploratory numerical experiments  should be reformulated and transformed into rigorous theorems in the near future.

Purposeful kinesis is possible even for very simple organisms: it requires only perception of local and instant information. For more complex organisms perception of non-local information, memory and prediction ability are possible and the kinesis should be combined with taxis.  The idea of evolutionary optimality can also be applied to taxis. This approach immediately produces an advection flux, which is  proportional to the gradient of the reproduction coefficient.  \cite{Cantrell2010} introduced and studied such models. Moreover,
the evolutionarily stable flux in these models should be proportional to $u\nabla r$  \citep{Averill2012, Gejji2012}. Here, $\nabla  r$ could be considered as a `driving force'. It would be a very interesting task to combine models of purposeful kinesis with these models of taxis and analyse the evolutionarily stable dispersal strategies, which are not necessarily  unique even for one species \citep{Buchi2012}. Of course, the {\em cost of mobility} should be subtracted from the reproduction coefficient for more detailed analysis.

If we go up the stair of organism complexity, more advanced effects should be taken into account like collective behaviour and interaction of groups in structured populations \citep{Perc2013}. Moreover, the evolutionary dynamics in more complex systems should not necessarily lead to an evolutionarily stable strategy and cycles are possible \citep{Gorban2007, Szolnoki2014}. Even relatively simple examples demonstrate that evolutionary dynamics can follow trajectories of an arbitrary dynamical system on the space of strategies \citep{Gorban(1984)}. Human behaviour can be modelled by differential equations with use of statistical physics and evolutionary games \citep{Perc2017}, but special care in model verifications and healthy scepticism in interpretation of the results are needed to avoid oversimplification.

\section*{Acknowledgement}We are very  grateful to the anonymous referees. Their enthusiastic and detailed comments helped us to improve the paper. AG and N\c{C} were supported by the University of Leicester, UK. AG was supported by the Ministry of education and science of Russia (Project No. 14.Y26.31.0022).


\begin{thebibliography}{99}

\bibitem[Allee et al.(1949)]{Allee1949}Allee, W.C., Park, O., Emerson, A.E., Park, T., Schmidt, K.P., 1949. Principles of animal ecology.  Saunders, Philadelphia.

\bibitem[Averill et al.(2012)]{Averill2012}Averill, I., Lou, Y., Munther, D., 2012. On several conjectures from evolution of dispersal. J. Biol. Dyn. 6(2), 117--130.

\bibitem[Gejji et al.(2012)]{Gejji2012}Gejji, R., Lou, Y., Munther, D., Peyton, J., 2012. Evolutionary convergence to ideal free dispersal strategies and coexistence.  Bull. Math. Biol.  74(2), 257--299.

\bibitem[Bazykin(1998)]{Bazykin1998}Bazykin, A.D., 1998. Nonlinear Dynamics of Interacting Populations. World Scientific, Singapore.

\bibitem[Buchi and Vuilleumier(2012)]{Buchi2012}B\"uchi, L., Vuilleumier, S., 2012. Dispersal strategies, few dominating or many coexisting: the effect of environmental spatial structure and multiple sources of mortality. PloS one, 7(4), p.e34733.

\bibitem[Cantrell et al.(2010)]{Cantrell2010}Cantrell, R.S., Cosner, C. and Lou, Y., 2010. Evolution of dispersal and the ideal free distribution. Mathematical biosciences and engineering: MBE. 7(1), 17--36.

\bibitem[Chalub et al.(2004)]{Chalub2004}Chalub, F.A., Markowich, P.A., Perthame, B., Schmeiser, C., 2004. Kinetic models for chemotaxis and their drift-diffusion limits. Monatshefte f\"ur Mathematik, 142(1--2), 123--141.

\bibitem[Chapman and Cowling(1970)]{Chapman1970}Chapman, S., Cowling, T.G., 1970. The mathematical theory of non-uniform gases: an account of the kinetic theory of viscosity, thermal conduction and diffusion in gases. Cambridge University Press.

\bibitem[Chen et al.(2010)]{Chen2010}Chen, W., Sun, H., Zhang, X. and Koro\v{s}ak, D., 2010. Anomalous diffusion modeling by fractal and fractional derivatives. Computers \& Mathematics with Applications, 59(5), pp.1754-1758.

\bibitem[Cosner(2014)]{Cosner2014}Cosner, C., 2014. Reaction-diffusion-advection models for the effects and evolution of dispersal. Discrete Contin. Dyn. Syst. 34, 1701--1745.

\bibitem[Dall et al.(2005)]{Dall2005}Dall, S.R., Giraldeau, L.A., Olsson, O., McNamara, J.M., Stephens, D.W., 2005. Information and its use by animals in evolutionary ecology. Trends Ecol. Evol. 20(4), 187--193.

\bibitem[Dobzhansky(1950)]{Dobzhansky1950}Dobzhansky, T., 1950. Evolution in the tropics,  Am. Sci. 38 (2), 209--221.

\bibitem[Gorban(1984)]{Gorban(1984)}Gorban A.N., 1984. Equilibrium Encircling. Nauka, Novosibirsk.

\bibitem[Gorban(2007)]{Gorban2007}Gorban, A.N., 2007. Selection theorem for systems with inheritance.
Math. Model. Nat. Phenom.  2(4), 1--45.

\bibitem[Gorban and Sadovskiy(1989)]{Gorban1989}Gorban, A.N., Sadovskiy, M.G., 1989. Optimal strategies of spatial distribution: the Allee effect. Zh. Obshch. Biol. 50(1), 16--21.

\bibitem[Gorban et al.(2016)]{Gorban2016}Gorban, A.N., Tyukina, T.A., Smirnova, E.V., Pokidysheva, L.I., 2016. Evolution of adaptation mechanisms: Adaptation energy, stress, and oscillating death. J. Theor. Biol. 405, 127--139.

\bibitem[Gr\"unbaum(1998)]{Grunbaum1998}Gr\"unbaum, D., 1998. Using spatially explicit models to characterize foraging performance in heterogeneous landscapes. The American Naturalist, 151(2), 97--113.

\bibitem[Gr\"unbaum(1999)]{Grunbaum1999}Gr\"unbaum, D., 1999. Advection–diffusion equations for generalized tactic searching behaviors. Journal of Mathematical Biology, 38(2), 169--194.

\bibitem[Hillen and Painter(2009)]{Hillen2009}Hillen, T., Painter, K.J., 2009. A user’s guide to PDE models for chemotaxis. J. Math. Biol. 58, 183--217.

\bibitem[Hofbauer and Sigmund(1998)]{Hofbauer1998}Hofbauer, J., Sigmund, K., 1998. Evolutionary games and population dynamics. Cambridge University Press.

\bibitem[Haldane(1932)]{Haldane1932} Haldane, J.B.S.,  1932. The Causes of Evolution,
    Longmans Green, London.

\bibitem[Keller and Segel(1971)]{Keller1971}Keller, E.F., Segel, L.A., 1971. Model for chemotaxis. J. Theor. Biol. 30(2), 225--234.

\bibitem[MacArthur and  Wilson(1967)]{MacArthurWilson1967}MacArthur, R.H.,  Wilson, E.0., 1967. The Theory  of  Island  Biogeography.  Princeton Univ. Press, Princeton,  N.J.

\bibitem[Maynard-Smith(1982)]{Maynard-Smith1982}Maynard-Smith, J. 1982. Evolution and the
    Theory of Games, Cambridge University Press, Cambridge.

 \bibitem[McCarthy(1997)]{McCarthy1997}McCarthy, M.A., 1997. The Allee effect, finding mates and theoretical models. Ecol. Model. 103(1), 99--102.

\bibitem[M{\'e}ndez et al.(2012)]{Mendez2012}M{\'e}ndez, V., Campos, D., Pagonabarraga, I., Fedotov, S., 2012. Density-dependent dispersal and population aggregation patterns. J. Theor. Biol. 309, 113--120.

\bibitem[M{\'e}ndez et al.(2010)]{Mendez2010}M{\'e}ndez, V., Fedotov, S. and Horsthemke, W., 2010. Reaction-transport systems: mesoscopic foundations, fronts, and spatial instabilities. Springer.

\bibitem[Metz et al.(1992)]{Metz1992}Metz, J.A., Nisbet, R.M., Geritz, S.A., 1992. How should we define ‘fitness’ for general ecological scenarios?. Trends. Ecol. Evol. 7(6), 198--202.

\bibitem[Morozov et al.(2006)]{Morozov2006}Morozov, A., Petrovskii, S., Li, B.L., 2006. Spatiotemporal complexity of patchy invasion in a predator-prey system with the Allee effect. J. Theor. Biol. 238(1), 18--35.

\bibitem[NDSolve(2014)]{NDSolve2014}NDSolve. 2014. Wolfram Language \& System Documentation Center.  url{http://reference.wolfram.com/language/ref/NDSolve.html}.

\bibitem[Nolting et al.(2015)]{Nolting2015}Nolting, B.C., Hinkelman, T.M., Brassil, C.E., Tenhumberg, B., 2015. Composite random search strategies based on non-directional sensory cues. Ecol. Complex. 22, 126--138.

\bibitem[Nonaka and Holme(2007)]{Nonaka2007}Nonaka, E., Holme, P., 2007. Agent--based model approach to optimal foraging in heterogeneous landscapes: effects of patch clumpiness. Ecography 30(6), 777--788.


\bibitem[Odum and Barrett(1971)]{Odum1971}Odum, E.P., Barrett, G.W., 1971. Fundamentals of ecology. Saunders, Philadelphia.

\bibitem[Othmer et al.(1988)]{Othmer1988}Othmer, H.G., Dunbar, S.R., Alt, W., 1988. Models of dispersal in biological systems. J. Math. Biol. 26(3), 263--298.

\bibitem[Othmer and Hillen(2000)]{Othmer2000}Othmer, H.G., Hillen, T., 2000. The diffusion limit of transport equations derived from velocity-jump processes. SIAM J. Appl. Math. 61(3), 751--775.

\bibitem[Othmer and Hillen(2002)]{Othmer2002}Othmer, H.G., Hillen, T., 2002. The diffusion limit of transport equations II: Chemotaxis equations. SIAM J. Appl. Math. 62(4), 1222--1250.

\bibitem[Parker and Smith(1990)]{Parker1990}Parker, G.A., Smith, J.M., 1990. Optimality theory in evolutionary biology. Nature 348(6296), 27--33.

\bibitem[Patlak(1953)]{Patlak1953}Patlak, C. S. 1953. Random walk with persistence and external bias. B. Math. Biophys. 15, 311--338.

\bibitem[pdpe(2017)]{pdpe2017}pdepe. 2017. Solve initial-boundary value problems for parabolic-elliptic PDEs in 1-D. MatWorks Documentation
\url{https://uk.mathworks.com/help/matlab/ref/pdepe.html}.

\bibitem[Perc et al.(2013)]{Perc2013}Perc, M., Gómez-Gardeñes, J., Szolnoki, A., Floría, L.M. and Moreno, Y., 2013. Evolutionary dynamics of group interactions on structured populations: a review. Journal of the royal society interface, 10(80), p.20120997.

\bibitem[Perc et al.(2017)]{Perc2017}Perc, M., Jordan, J.J., Rand, D.G., Wang, Z., Boccaletti, S. and Szolnoki, A., 2017. Statistical physics of human cooperation. Physics Reports,  687, 1--51.


\bibitem[Petrovskii et al.(2002)]{Petrovskii2002}Petrovskii, S.V., Morozov, A.Y., Venturino, E., 2002. Allee effect makes possible patchy invasion in a predator--prey system. Ecol. Lett. 5(3), 345--352.

\bibitem[Pianka(1970)]{Pianka1970}Pianka, E.R., 1970. On $r$- and $K$-selection, Am. Nat. 104 (940), 592--597.

\bibitem[Pough(1989)]{Pough1989}Pough, F.H., 1989. Performance and Darwinian Fitness: Approaches and Interpretations, Physiol. Zool.  62 (2), 199--236.

 \bibitem[Rahmandad and Sterman(2008)]{Rahmandad2008}Rahmandad, H., Sterman, J., 2008. Heterogeneity and network structure in the dynamics of diffusion: Comparing agent-based and differential equation models. Manage. Sci. 54(5), 998--1014.

\bibitem[Rosenblueth and Wiener(1950)]{Rosenblueth1950}Rosenblueth, A., Wiener, N., 1950. Purposeful and non-purposeful behavior. Philos. Sci. 17(4), 318--326.

\bibitem[Sadovskiy et al.(2009)]{Sadovskiy2009}Sadovskiy, M.G., Senashova, M.Y., Brychev, P.A., 2009. A model of optimal migration of locally informed beings.  Dokl. Math.  80(1), 627--629.

\bibitem[Szolnoki et al.(2014)]{Szolnoki2014}Szolnoki, A., Mobilia, M., Jiang, L.L., Szczesny, B., Rucklidge, A.M. and Perc, M., 2014. Cyclic dominance in evolutionary games: a review. Journal of the Royal Society Interface, 11(100), p.20140735.


\bibitem[Turchin(1989)]{Turchin1989}Turchin, P., 1989. Population consequences of aggregative movement. J. Anim. Ecol. 58(1), 75--100.

\bibitem[Tyutyunov et al.(2017)]{Tyutyunov2017}Tyutyunov, Y.V., Titova, L.I., Senina, I.N., 2017. Prey-taxis destabilizes homogeneous stationary state in spatial Gause--Kolmogorov--type model for predator--prey system. Ecol. Complex. 31, 170--180.


\bibitem[Wainwright(1994)]{Wainwright1994}Wainwright, P.C., 1994. Functional morphology as a tool for ecological research. In P.C. Wainwright, S.M. Reilly (eds.), Ecological morphology: integrative organismal biology. University of Chicago Press, Chicago, 42--59.

\end{thebibliography}
\end{document}